\documentclass[%
 reprint,
superscriptaddress,
 amsmath,amssymb,
]{revtex4-2}

\setcitestyle{super}

\usepackage{graphicx}
\usepackage{dcolumn}
\usepackage{bm}
\usepackage{hyperref}
\usepackage[capitalise]{cleveref}
\usepackage{physics}
\usepackage{braket}
\usepackage{xcolor}
\usepackage{stmaryrd}

\newcommand{\PsiT}{\Psi_\text{T}}
\newcommand{\pol}{\boldsymbol{\epsilon}}

\newcommand{\Qph}{Q_\text{ph}}
\newcommand{\Piph}{\Pi_\text{ph}}
\newcommand{\Vext}{V_\text{ext}}
\newcommand{\dg}{{}^\dag}
\newcommand{\phdg}{^{\vphantom{\dag}}}
\newcommand{\Helel}{H_2}
\newcommand{\Helph}{H_{1}}
\newcommand{\Hph}{H_\text{ph}}
\newcommand{\Ec}{E_\text{c}}

\begin{document}
\title{Phaseless auxiliary-field quantum Monte Carlo method\\for cavity-QED matter systems}
\author{Lukas Weber}
\email{lweber@flatironinstitute.org}
\affiliation{Center for Computational Quantum Physics, The Flatiron Institute, 162 Fifth Avenue, New York, NY 10010, USA}
\affiliation{Max Planck Institute for the Structure and Dynamics of Matter, Luruper Chaussee 149, 22761 Hamburg, Germany}
\author{Leonardo dos Anjos Cunha}
\affiliation{Center for Computational Quantum Physics, The Flatiron Institute, 162 Fifth Avenue, New York, NY 10010, USA}
\author{Miguel A. Morales}
\affiliation{Center for Computational Quantum Physics, The Flatiron Institute, 162 Fifth Avenue, New York, NY 10010, USA}
\author{Angel Rubio}
\affiliation{Max Planck Institute for the Structure and Dynamics of Matter, Luruper Chaussee 149, 22761 Hamburg, Germany}
\affiliation{Center for Computational Quantum Physics, The Flatiron Institute, 162 Fifth Avenue, New York, NY 10010, USA}
\author{Shiwei Zhang}
\affiliation{Center for Computational Quantum Physics, The Flatiron Institute, 162 Fifth Avenue, New York, NY 10010, USA}

\begin{abstract}
\begin{centering}
    \hfill \textbf{Abstract} \hfill\medskip
\end{centering}

We present a generalization of the phaseless auxiliary-field quantum Monte Carlo (AFQMC) method to cavity quantum-electrodynamical (QED) matter systems. The method can be formulated in both the Coulomb and the dipole gauge. We verify its accuracy by benchmarking calculations on a set of small molecules against full configuration interaction and state-of-the-art QED coupled cluster (QED-CCSD) calculations. Our results show that (i) gauge invariance can be achieved within correlation-consistent Gaussian basis sets, (ii) the accuracy of QED-CCSD can be enhanced significantly by adding the standard perturbative triples correction without light-matter coupling, and (iii) there is a straightforward way to evaluate the differential expression for the photon occupation number that works in any gauge.
The high accuracy and 
favorable computational scaling of our AFQMC approach will enable a broad range of applications. 
Besides polaritonic chemistry, the method opens a way to simulate extended QED matter systems.
\end{abstract}
\maketitle
\section{Introduction}
The enhanced light-matter coupling in electromagnetic cavities can turn the usually weak quantum fluctuations of light into a driver of change in materials and chemistry, that promises to achieve light-driven control without the heating problem usually associated with external laser driving.
This potential has created considerable excitement in various applications and generated intense research activities from both physics and chemistry.

For instance, several experimental groups have reported considerable differences in chemical reactivity and molecular properties when performing experiments outside and inside an optical cavity resonant to certain vibrational and/or electronic excitations of the matter system.\cite{thomas2019tilting,ahn2023modification,hutchison2012m,sandeep2022manipulating,joseph2021supramolecular,joseph2024consequences,patrahau2024direct}
For condensed matter systems, it has been shown how coupling to a cavity can shift thermal phase transitions \cite{jarc_cavitymediated_2023} or lead to a breakdown of the quantum Hall effect.\cite{appugliese_breakdown_2022}

These cavity quantum systems, apart from posing the experimental challenge of designing efficient cavities and coupling them to matter effectively, have posed a significant challenge to the development of numerical methods that can treat matter and photons on the same footing. On the one hand, this has given rise to ab-initio methods such as the quantum electrodynamical density functional theory (QEDFT)~\cite{ruggenthaler_quantumelectrodynamical_2014,tokatly_timedependent_2013,pellegrini_optimized_2015,schafer_making_2021,tasci_photon_2024,lu_electronphoton_2024}. On the other hand, major efforts have been made to adapt precision methods for more correlated systems 
to the cavity QED setting, including DMRG~\cite{eckhardt_quantum_2022,passetti_cavity_2023,shaffer_entanglement_2024,matousek_polaritonic_2024}, coupled cluster methods (CCSD)~\cite{haugland_coupled_2020,pavosevic_polaritonic_2021,pavosevic_computational_2023}, complete active space configuration interaction~\cite{vu_cavity_2023}, and different flavors of quantum Monte Carlo, such as the stochastic series expansion (SSE)~\cite{weber_quantum_2022,weber_cavityrenormalized_2023,langheld_quantum_2024}, and diffusion Monte Carlo (DMC)~\cite{weight_diffusion_2024}.
Given the inherent challenges of accurate 
treatment of electronic correlation effects,
the goal of treating  photons together while maintaining systematic accuracy is clearly a daunting task.
It is not surprising that significant challenges still exist in each of these approaches, and a systematic and general many-body computational method for cavity QED is still lacking.

In this work, we will present a quantum electrodynamical auxiliary-field quantum Monte Carlo (QED-AFQMC) method.  
Compared to the SSE, which is limited to bosonic lattice models, and DMC, which is most accurate in systems with spatially separated electrons, AFQMC~\cite{zhang_constrained_1997,zhang_quantum_2003} has proven to be versatile in its accuracy in various situations, such as in Hubbard-like models~\cite{leblanc_solutions_2015,qin_absence_2020,xiao_temperature_2023,xu_coexistence_2024}, the realm of quantum chemistry~\cite{williams_direct_2020,shee-transformative-jcp-2023,motta_initio_2018,lee_twenty_2022}, and the realistic description of materials~\cite{purwanto_pressureinduced_2009,ma_quantum_2015,chen_initio_2021}. Techniques such as the constrained path~\cite{zhang_constrained_1997} and the phaseless approximation~\cite{zhang_quantum_2003} allow, given an appropriate trial wavefunction, to compute an accurate solution even in the presence of a sign or phase problem.

Building on earlier work of applying AFQMC to electron-phonon lattice models~\cite{lee_constrainedpath_2021}, we show how to perform AFQMC calculations on a generic Pauli-Fierz Hamiltonian describing matter in a single mode cavity, in both the Coulomb and the dipole gauges. We then apply the method to a set of small molecules, comparing to QED-CCSD-2 calculations\cite{pavosevic2022cavity}. While different gauges are known to only agree in the complete basis-set limit, we demonstrate 
that this convergence can be achieved in the Gaussian basis-set framework. Finally, we provide a conclusion and outlook for the further development of the method and potential applications.

\section{Model}
Within the Born-Oppenheimer approximation, we consider a collection of electrons subject to an external potential $\Vext$, the longitudinal Coulomb interaction, and a coupling to a single mode in the long-wavelength approximation. In the Coulomb gauge and atomic units, the corresponding Pauli-Fierz Hamiltonian can be written as
\begin{align}
\label{eq:first_quant}
H^\text{C} &= \Helph^\text{C} + \Helel^\text{C} + \Hph
\end{align}
with
\begin{align}
\Helph^\text{C} &= \sum_{i} \frac 12 (\mathbf{p}_i + \mathbf{A})^2 + \Vext(\mathbf{r}_i),\nonumber\\
\Helel^\text{C} &= \frac 12 \sum_{i\neq j} \frac{1}{|\mathbf{r}_i-\mathbf{r}_j|},\nonumber\\
\Hph &= \frac \Omega 2 (\Piph^2 + \Qph^2 - 1),
\end{align}
where in $\Hph$
the photon displacement and momentum operators are given by $\Qph$ and $\Piph$, respectively,
and $\mathbf{A} = \Qph \pol/\sqrt{\Omega}$ in $\Helph^\text{C}$. The unnormalized polarization vector $\pol$ encodes the strength of the cavity electric field at the position of the matter system, and $\Omega$ is the cavity frequency.
Alternatively, via the Power-Zienau-Woolley transformation $U_\text{PZW} = \exp(-i \sum_k \mathbf r_k \cdot \mathbf{A})$ and a canonical transformation $\Piph \leftrightarrow \Qph$, we can write the Hamiltonian in the dipole gauge as
\begin{align}
\label{eq:dipolegauge}
H^\text{D} &= \Helph^\text{D} + \Helel^\text{D} + \Hph,\nonumber\\
\Helph^\text{D} &= \sum_{i} \frac 12 \mathbf{p}_i^2 + \sqrt{\Omega} \mathbf r_i \cdot \pol\, \Qph + \Vext(\mathbf{r}_i),\nonumber\\
\Helel^\text{D} &= \frac 12 \sum_{i\neq j} \frac{1}{|\mathbf{r}_i-\mathbf{r}_j|} + \frac 12 \qty(\pol \cdot \sum_i \mathbf{r}_i)^2.
\end{align}

While the two gauges are related by a unitary transformation in the complete basis set limit, the dipole gauge has been shown to be more accurate under basis set truncation~\cite{debernardis_breakdown_2018,li_electromagnetic_2020}. We further improve the convergence to the gauge-invariant limit by writing the dipole-dipole interaction as the square of basis-truncated dipoles, rather than the exact quadrupole matrix elements~\cite{taylor_resolution_2020} (see \cref{app:quadrupole} for a comparison).

By contrast, in periodic systems, the dipole moments in $H^\text{D}$ break translational symmetry of the electronic sector\footnote{In the complete Hilbert space, translation symmetry is restored by simultaneously displacing $\Qph$.}, whereas the Coulomb gauge expression is manifestly translation invariant and can be conveniently formulated in a plane-wave basis.

The AFQMC method presented in the following can work in both gauges and thus profits from their unique benefits.
Additionally it
provides
a way to check results for gauge invariance,
which we also demonstrate here.

\section{Method}
One way to obtain the ground state of the Hamiltonian $H$
is to perform an imaginary time projection,
\begin{align}
\label{eq:timeprop}
\ket{\Psi_0} = \lim_{\beta\to\infty} e^{-\beta (H-E_0)} 
\ket{\Psi_\text{I}}\,,
\end{align}
where $\ket{\Psi_\text{I}}$ is an initial wave function which is not orthogonal to $\ket{\Psi_0}$.
The basic idea of the AFQMC method is to express this many-body time evolution as an integral over one-body time evolutions, which individually can be evaluated at low cost. To achieve this mapping, we first need to rewrite the Hamiltonian in terms of one-body operators.

\subsection{Rewriting the Hamiltonian}
The first step towards this goal is to rewrite the electronic part of the Hamiltonian in the second quantization, which for both $H^\text{C}$ and $H^\text{D}$ can be expressed as
\begin{align}
\label{eq:H-2nd-quant}
H &= \sum_{ij,\sigma} h_{ij}(\Qph)\, c\dg_{i\sigma} c\phdg_{j\sigma}\nonumber\\ &+ \frac 1 2 \sum_{ijkl,\sigma \sigma'} v_{ijkl}\,c\dg_{i\sigma} c\dg_{j\sigma'} c\phdg_{l\sigma'} c\phdg_{k\sigma}\nonumber\\
&+ \frac \Omega 2 (\Piph^2 + \Qph^2 - 1)\,.
\end{align}
Here, the light-matter coupling gives rise to a functional dependence of $h_{ij}(\Qph) = \braket{i|\Helph^\text{C/D}|j}$ on the photon displacement $\Qph$.

The electron repulsion integrals in Eq.~(\ref{eq:H-2nd-quant}) also acquire gauge dependence:
$v_{ijkl} = \braket{ij|\Helel^\text{C/D}|kl}$.
In general, it is possible to decompose
it into a sum of squares $v_{ijkl} = \sum_\gamma \nu_{ik}^\gamma \nu_{jl}^\gamma$ that allows rewriting the Hamiltonian into the form
\begin{align}
H &= \sum_{\substack{ij\\\sigma}} \qty[h_{ij}(\Qph) - \frac 12 \sum_{k} v_{ikkj}]\, c\dg_{i\sigma} c\phdg_{j\sigma}\nonumber\\ &+ \frac 1 2 \sum_{\gamma} \qty(\sum_{\substack{ij,\sigma}} \nu^\gamma_{ij} \,c\dg_{i\sigma} c\phdg_{j\sigma})^2+ \Hph\,.
\end{align}
The decomposition above can be calculated numerically, e.g., using the modified Cholesky decomposition~\cite{purwanto_assessing_2011} or interpolative separable density fitting~\cite{malone_overcoming_2019}. In the plane-wave basis, a decomposition can also be written down analytically in terms of density operators~\cite{suewattana_phaseless_2007}. 
The number of terms in the sum of squares, $\gamma$, is of ${\mathcal O}(N)$, where 
$N$ is the size of the single-particle basis (the space for $i$,$j$, $k$,$l$).
When working in the dipole gauge, the additional term in $H_2^\text{D}$ from \cref{eq:dipolegauge} is already squared and leads to just one additional operator $\nu_0 = \pol\cdot \mathbf r$. If the exact quadrupole matrix elements are not omitted (see \cref{app:quadrupole}), another single particle piece appears that can be absorbed into $h_{ij}(\Qph)$.

In practical calculations, it is beneficial to subtract the mean-field background~\cite{purwanto_pressureinduced_2009} $\bar{\nu}_\gamma = \braket{\PsiT|\nu_\gamma|\PsiT}$ (for all $\gamma$) from the interaction operators 
before proceeding. Defining $\vec{\nu}$ as the vector containing the operators $\nu_\gamma = \sum_{ij\sigma} \nu^\gamma_{ij} c\dg_{i\sigma} c\phdg_{j\sigma}$, we can rewrite the electron-electron interaction part of $H$ in terms of new operators $v_\gamma = \nu_\gamma - \bar{\nu}_\gamma$ as 
\begin{align}
\label{eq:Hsquared}
H &= \Helph + \Helel + \Hph\nonumber\\
\Helph &= \sum_{\substack{ij\\\sigma}} \qty[h_{ij}(\Qph) - \frac 12 \sum_{k} v_{ikkj}]\, c\dg_{i\sigma} c\phdg_{j\sigma}+ \vec{v}\cdot \vec{\bar{\nu}} + \frac{\vec{\bar{\nu}}^2}{2}\nonumber\\
\Helel &= \frac{\vec{v}^2}{2}.
\end{align}

The form of \cref{eq:Hsquared} allows us to Suzuki-Trotter decompose the imaginary time propagator for small $\Delta\tau$ as
\begin{align}
\label{eq:trotter}
e^{-\Delta\tau H} &= e^{-\frac{\Delta\tau}{2} \Hph} e^{-\frac{\Delta\tau}{2} {\Helph}} e^{-\Delta\tau {\Helel}} e^{-\frac{\Delta\tau}{2} {\Helph}}e^{-\frac{\Delta\tau}{2} \Hph}\nonumber\\
&+ \mathcal{O}(\Delta\tau^3)
\end{align}
and then apply the Hubbard-Stratonovich transformation
\begin{align}
e^{-\Delta\tau \Helel} = e^{-\frac{\Delta\tau}{2} \vec{v}^2} = \int_{-\infty}^\infty \frac{d\vec{x}}{\sqrt{2\pi}} e^{-\frac{1}{2}\vec{x}^2} e^{i\sqrt{\Delta\tau} \vec{x} \cdot \vec{v}},
\end{align}

For the propagation of $\Hph$, we employ a similar technique as Ref.~\onlinecite{lee_constrainedpath_2021}, but write the harmonic propagator exactly
\begin{align}
\label{eq:SHO-prop}
&\braket{q'|e^{-\Delta\tau \Hph}|q} = \frac{1}{\sqrt{2\pi \sinh(\Omega\Delta\tau)}}\nonumber\\
&\times \exp[-\frac{\qty(q'-\frac{q}{\cosh(\Omega\Delta\tau)})^2}{2\tanh(\Omega \Delta\tau)}- \frac{\tanh(\Omega\Delta\tau)q^2-\Omega\Delta\tau}{2}],
\end{align}
which is more accurate when $\Omega$ is larger than the relevant electronic energy scale, without loss of efficiency. Note, for example, that in the limit $\Omega\to\infty$, $q'$ is projected to the photon ground state in every step, which reproduces the result of high-frequency perturbation theory $H_\text{eff} =\braket{0_\text{ph}|H|0_\text{ph}} + \mathcal{O}(1/\Omega)$.
\subsection{Importance sampling}
The building blocks described above allow us to rewrite the time-projected wavefunction as a weighted average of light-matter product states
\begin{align}
\label{eq:walkers}
\ket{\Psi(\tau)} 
= e^{-\tau H} 
\ket{\Psi_\text{I}} 
\approx \sum_n w_n(\tau) \ket{\Phi_{\text{el},n}(\tau), q_n(\tau)},
\end{align}
where the weights $w_n$, the Slater determinants $\Phi_{\text{el},n}$, and the photon displacements $q_n$ follow a stochastic process that implements the repeated action of $e^{-\Delta \tau H}$. For the initial state $\ket{\Psi_\text{I}} = \sum_n \ket{\Phi_{\text{el},n}(0), q_n(0)}$, we sample a population of walkers according to their overlap with the trial wavefunction $\ket{\PsiT}$.

After an importance sampling transformation with respect to a trial wavefunction $\ket{\PsiT}$, \cref{eq:walkers} becomes
\begin{align}
\ket{\Psi(\tau)}
&\approx \sum_n w_n(\tau) \frac{\ket{\Phi_{\text{el},n}(\tau), q_n(\tau)}}{\braket{\PsiT|\Phi_{\text{el},n}(\tau), q_n(\tau)}}\,,
\end{align}
with the adjusted stochastic process
\begin{equation}
\ket{\Phi_\text{el}',q'} 
\mapsfrom
\frac{\braket{\PsiT|\Phi_\text{el}',q'}}{\braket{\PsiT|\Phi_\text{el},q}} e^{-\Delta \tau H}\ket{\Phi_\text{el},q},
\end{equation}
where the primed quantities are at step $\tau+\Delta\tau$ and the unprimed ones are at step $\tau$.

In the following, we describe this process for each of the components of the propagator in \cref{eq:trotter}. We note that apart from the fermionic subsystem, the algorithm is similar to the previously derived AFQMC for lattice phonon systems~\cite{lee_constrainedpath_2021}.

\subsubsection{One-body propagator}
First, the application of 
the electronic one-body part,
$e^{-\frac{\Delta\tau}{2} H_1(\Qph)}$, leaves $q$ invariant, since $\ket{\Phi_{\text{el}}(\tau), q(\tau)}$ is an eigenstate of $\Qph$,  leading to
\begin{align}
q' &= q,\nonumber\\
\Phi'_\text{el} 
&= e^{-\frac{\Delta\tau}{2} \Helph}(q) 
\Phi_\text{el},
\nonumber\\
w' &= \frac{\braket{\PsiT|\Phi_{\text{el}}', q'
}}{\braket{\PsiT|\Phi_{\text{el}}, q}} w(\tau).
\end{align}

\subsubsection{Two-body propagator}
Similarly, in the second part, the application of $e^{-\Delta\tau H_2}$ remains standard. We sample the auxiliary fields $\vec{x}$ from a normal distribution that is shifted by the force-bias $\vec{f} = -\braket{\PsiT| i\sqrt{\Delta\tau}
\vec{v}|\Phi_\text{el}, q}$. 
This shift has to be canceled out in the weight:
\begin{align}
q' &= q,\nonumber\\
\Phi_\text{el}' &= e^{i\sqrt{\Delta\tau}(\vec{x}-\vec{f})\cdot \vec{v}} \Phi_\text{el},\nonumber\\
w' &= \frac{\braket{\PsiT|\Phi_{\text{el}}', q'
}}{\braket{\PsiT|\Phi_{\text{el}}, q}} e^{\vec{x}\cdot \vec{f}-\frac{\vec{f}^2}{2}} w.
\end{align}

\subsubsection{Photon propagator}
Third is the application of the harmonic propagator. We write
\begin{align}
\label{eq:approxprop}
&\frac{\braket{\PsiT|\Phi_{\text{el}}', q'}}{\braket{\PsiT|\Phi_{\text{el}}, q}} e^{-\frac{\Delta\tau}{2} H_\text{ph}} \ket{\Phi_\text{el}, q}\nonumber \\
&\approx \int dq' e^{(q'-q) \partial_q\log \PsiT(q)} \ket{\Phi_\text{el}, q'}\!\braket{q'|e^{-\frac{\Delta\tau}{2}H_\text{ph}}|q}
\end{align}
where $\PsiT(q) = \braket{\PsiT|\Phi_\text{el}, q}$, 
and sample $q'$ from a normal distribution
\begin{align}
q' &= \mathcal{N}\qty(\frac{q}{\cosh{\delta}} + \tanh{\delta} \frac{\partial_q\PsiT(q)}{\PsiT(q)}, \tanh{\delta})\nonumber\\
\Phi_\text{el}' &= \Phi_\text{el}\nonumber\\
w' &= \frac{\braket{\PsiT|\Phi_{\text{el}}', q'}}{\braket{\PsiT|\Phi_{\text{el}}, q}} e^{(\frac{q}{\cosh{\delta}}-q')\partial_q\log \PsiT(q)} \frac{1}{\sqrt{\cosh\delta}}\nonumber\\
 &\times e^{- \frac{\tanh\delta}{2} \qty[-(\partial_q \log\PsiT(q))^2 + q^2] + \frac{\delta}{2}} w,
\end{align}
where $\delta = \Omega\Delta\tau/2$.
(Note that the propagator 
in \cref{eq:approxprop} is for 
a ``half'' time step $\Delta\tau/2$,
while the general formula in 
\cref{eq:SHO-prop} is written for $\Delta\tau$.)
In the weight update, we include the true weight ratio to cancel out the approximation that was made in \cref{eq:approxprop}.

While we concentrated on a single mode here, the generalization to multiple modes should not make the method much more costly, and in fact has already been performed in the case of phonons on the lattice~\cite{lee_constrainedpath_2021}. However, as long as the long-wavelength approximation is not relaxed, some multimode systems can be mapped back to an effective single-mode model~\cite{svendsen_theory_2023}.

\subsection{Phaseless constraint}
In the stochastic process defined above, the weights $w_n(\tau)$ are complex numbers and can be subject to phase fluctuations that generally arise from both the electron-electron interactions and the light-matter interactions.\footnote{A sign- and phase-problem-free simulation can be obtained (i) in the absence of the Coulomb repulsion, (ii) with inversion symmetry along the polarization direction, and (iii) zero magnetization.}

If phase fluctuations are present, they will lead to cancellations of the weights and prohibitively large statistical errors, also known as the phase problem~\cite{zhang_initio_2018}.
The AFQMC method relies on 
a constraint~\cite{zhang_constrained_1997,zhang_quantum_2003} on the random walk 
derived from an exact condition 
in the limit of an exact $\ket{\PsiT}$.
Part of this is already achieved by the importance sampling transformation and the force bias introduced above, which guide the random walk away from regions where $\braket{\PsiT|\Phi(\tau)} = 0$. The 
phaseless approximation~\cite{zhang_quantum_2003} then relies on the resulting imaginary force-bias $\vec{f}$ to achieve a constant gauge, which is preserved in the random walk via the projection
\begin{align}
 w' = |w| \max{\qty{0, \cos(\arg \frac{\braket{\PsiT|\Phi'_\text{el}, q'}}{\braket{\PsiT|\Phi_\text{el}, q}})}}
\end{align}
that suppresses large phase fluctuations that would wind around the origin$\braket{\PsiT|\Phi(\tau)} = 0$ and lead to cancellations. With an appropriate $\ket{\PsiT}$, the result is approximate. The bias decreases 
as $\ket{\PsiT}$ is improved, and it
has been shown to be small in a variety of systems using readily accessible trial wavefunctions.~\cite{qin_benchmark_2016,lee_twenty_2022}

\subsection{Trial wavefunction}
For our trial wavefunction, we will employ a simple product state of an unrestricted Hartree-Fock (UHF) Slater determinant and a squeezed coherent state.
\begin{equation}
\label{eq:trial}
\ket{\PsiT} = 
\ket{\Phi_\text{UHF}}
\otimes \int dq \sqrt[4]{\frac s \pi} e^{- \frac{s (q-q_0)^2}{2}}.
\end{equation}
where we variationally minimize the energy with respect to the orbital degrees of freedom of
$\ket{\Phi_\text{UHF}}$ 
and the parameters $s$ and $q_0$.
In the dipole gauge, $s=1$ and $q_0$ can be finite depending on the dipole moment of the system. By contrast, in the Coulomb gauge, $q_0$ typically (if $\pol(\vb r)$ does not describe a magnetic field \cite{bacciconi_firstorder_2023}) vanishes and $s=\sqrt{1+N\pol^2/\Omega^2}$.

In cases where $q_0$ is not known, there can be local minima in the energy landscape. In such cases -- albeit they do not appear in the benchmark systems of this paper -- we have found it useful to minimize all parameters numerically e.g. using the L-BFGS algorithm~\cite{liu_limited_1989} with exact energy gradients obtained from automatic differentiation.
\subsection{Population control and stabilization}
To stabilize the method, several further technical steps \cite{zhang_constrained_1997,motta_initio_2018} must be implemented. Firstly, in a population control step, we use the simple comb algorithm~\cite{booth_weight_1996} to stochastically eliminate walkers with small weights and replicate those with large weights in a way that approximately preserves the total distribution up to a small bias that vanishes for large walker populations. 

Second, to preserve the orthogonality of the Slater determinants, we reorthogonalize them regularly using a QR decomposition.

Finally, we cap the force bias and weight change as described in Ref.~\onlinecite{motta_initio_2018}. These caps help control rare events in the sampling but become increasingly loose when decreasing the timestep $\Delta\tau$.
\section{Benchmark results}
In this section, we apply our method to compute the properties of small cavity-coupled molecules in their equilibrium geometry. To this end, we always assume a single cavity mode of frequency $\Omega = 0.3\,\text{Ha}$ that is polarized along a symmetry axis of the molecule. The correlation consistent bases (aug)-cc-pVXZ from the Dunning family~\cite{dunning1989gaussian,kendall1992electron} were employed in the computations reported in this work. As usual, X refers to the cardinality of the basis set, and we limited our analysis to X=D,T,Q for double, triple and quadrupole-zeta bases, respectively.

\subsection{H\textsubscript{2}}
\begin{figure}
\begin{center}
\includegraphics{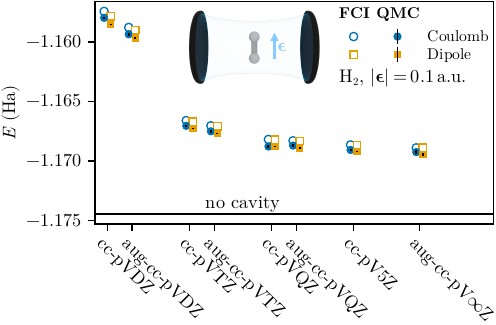}
\end{center}
\caption{The ground state energy, $E$, of the H\textsubscript{2} molecule coupled to a cavity mode, in different basis sets and gauges, computed with AFQMC and FCI. In the FCI calculation, the photonic Hilbert space is truncated to $n_\text{ph} < 30$. For comparison, the horizontal line shows the FCI ground state energy in the complete basis set limit without a cavity. The complete basis set limit (aug-cc-pV$\infty$Z) is obtained using cubic extrapolation of the aug-cc-pVTZ and aug-cc-pVQZ bases. The inset shows the orientation of the cavity mode polarization with respect to the molecule.}
\label{fig:h2_gauges}
\end{figure}
First, let us consider the case of H\textsubscript{2} in a cavity, where we can compare AFQMC calculations in both the Coulomb and the dipole gauge to full configuration interaction (FCI) calculations up to large basis set sizes, as shown in~\cref{fig:h2_gauges}. We see that in general, there is a good agreement between AFQMC and FCI, with the former slightly undershooting by about $0.5\,\text{mHa}$ on average. This error stems from the phaseless constraint of AFQMC and can likely be improved by choosing a more elaborate trial wavefunction that already incorporates some of the electron-photon correlations.

Furthermore, with increasing basis set size, the difference between the two gauges vanishes. After a cubic extrapolation to the complete basis set limit based on the augmented triples-zeta and quadruples-zeta bases~\cite{halkier_basisset_1998}, we find that the energy differences between the bases are reduced to $E^D_\text{FCI} - E^C_\text{FCI}=7\,\mu\text{Ha}$ and $E^D_\text{QMC} - E^C_\text{QMC} = -0.2(2)\text{mHa}$.

\subsection{HF, H\textsubscript{2}O, NH\textsubscript{3}, and CH\textsubscript{4}}
\begin{figure}
\begin{center}
\includegraphics{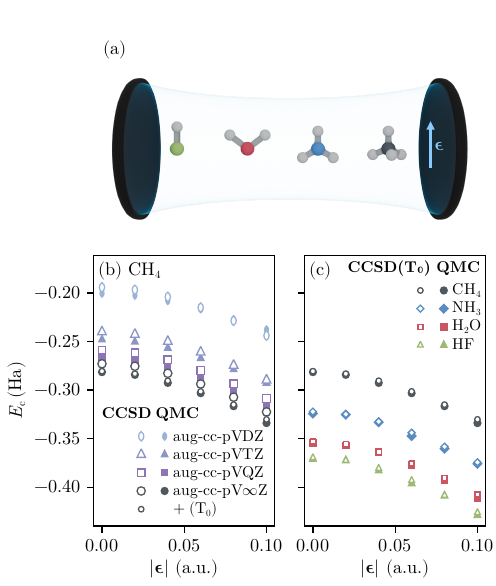}
\end{center}
\caption{The total correlation energy, $\Ec$, as a function of the light-matter coupling $|\pol|$, of the molecules HF, H\textsubscript{2}O, NH\textsubscript{3}, and CH\textsubscript{4}. (a) Sketch of the molecule orientation with respect to the cavity polarization $\pol$. (b) Scaling with basis set size for CH\textsubscript{4}. AFQMC energies are in the Coulomb gauge and CCSD energies in the dipole gauge, respectively. The final comparison requires an extrapolation to the complete basis set limit (black circles). The small open symbols show CCSD including the standard perturbative triples correction 
 obtained
without light-matter interactions. (c) The correlation energies extrapolated to the complete basis set limit for the whole range of molecules. The QMC statistical errorbars are smaller than the symbols.}
\label{fig:mols}
\end{figure}

Next, we compare our AFQMC results for the molecules HF, H\textsubscript{2}O, NH\textsubscript{3}, and CH\textsubscript{4} (\cref{fig:mols}(a)) to QED-CCSD-2 calculations (see \cref{app:ccsd1} for a comparison to QED-CCSD-1). While the CC calculations are performed in the dipole gauge, we perform the AFQMC calculations in Coulomb gauge. For both, we calculate the correlation energy as
\begin{equation}
\Ec = E - E^\text{C}_\text{Ref},
\end{equation}
where $E^\text{C}_\text{Ref} = \braket{\PsiT|H^C|\PsiT}$ with the product-state trial wavefunction from \cref{eq:trial} in the Coulomb gauge. Since the matter part of $\ket{\PsiT}$ is a single Slater determinant, $\Ec$ measures both the correlation energy due to the electron-electron and the electron-photon interactions. It is important to note that the reference energy is not gauge invariant and calculating it in the dipole gauge would already include some of the light-matter correlations due to the nature of the PZW transformation. 

As the two datasets in \cref{fig:mols}(b) belong to different gauges, we observe some differences at finite basis set sizes. Most notably, in the doubles-zeta basis set, there is a crossing between the CCSD and QMC correlation energies. This can be understood as the crossover of two sources of inaccuracy: The less accurate description of the pure electronic correlations by CCSD and the basis-set error of the Coulomb gauge, which increasingly affects the QMC data at large $|\pol|$.
To remove the latter, we therefore first need to extrapolate to the complete basis set limit, which we do using the same a cubic extrapolation scheme as in the case of H\textsubscript{2}.

After extrapolation, there is still an average gap of around $11\,\text{mHa}$ between the CC and AFQMC correlation energies (\cref{fig:mols}(b)), which can be significantly reduced to 1.7$\,$mHa by adding the purely electronic perturbative triples correction (T\textsubscript{0}) to CCSD, which we dub CCSD(T\textsubscript{0}). In contrast to the full light-matter triples correction, which, so far, has not been derived, the readily available (T\textsubscript{0}) correction is already enough to reach good agreement between the methods (\cref{fig:mols}(c)). In \Cref{app:nonint}, we further motivate this modification by comparing the methods in absence of the Coulomb repulsion.

\begin{figure}
\begin{center}
\includegraphics{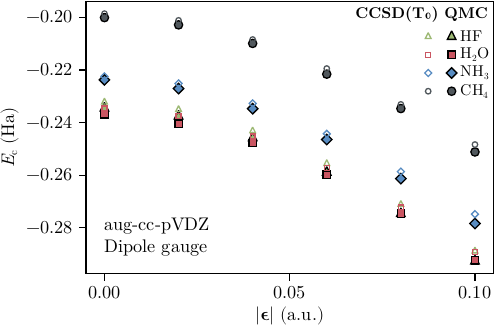}
\end{center}
\caption{The total correlation energy, $\Ec$, as a function of the light-matter coupling for the molecules HF, H\textsubscript{2}, NH\textsubscript{3}, CH\textsubscript{4}, computed by CCSD(T\textsubscript{0}) and AFQMC, both in the dipole gauge, in the aug-cc-pVDZ basis set. The QMC statistical errorbars are smaller than the symbols.}
\label{fig:dipole_fixed_basis}
\end{figure}
To compare to CCSD(T\textsubscript{0}) data without extrapolation, we can further perform AFQMC calculations directly in the dipole gauge in a fixed basis (\cref{fig:dipole_fixed_basis}). We observe a similar level of agreement in this case, albeit minimally worse than the comparison to the Coulomb gauge in the complete basis set limit. This could point to a stronger phaseless constraint error in the dipole gauge. Both \cref{fig:mols} and \cref{fig:dipole_fixed_basis} show a weak growth of the deviation with $|\pol|$, which can likely be attributed to a combination of phaseless constraint error and the missing $|\pol|$ dependence in T\textsubscript{0}.
The constraint error can be reduced by use of a multi-determinant $\ket{\Psi_T}$ as has been widely done in standard quantum chemistry applications, as well as improved electron-photon correlations. However these results are sufficient for the purpose of introducing our method and benchmarking, and
we leave the study of improved 
$\ket{\Psi_T}$
for a future investigation.

\subsection{Photon number of HF}
Finally, we calculate the photon occupation number $n_\text{ph} = \braket{\Piph^2+\Qph^2-1}/2$. Care has to be taken when computing this quantity in the dipole gauge, as the $U_\text{PZW}$ and the subsequent canonical transformation have changed the meaning of $\Qph$ and $\Piph$~\cite{foley_initio_2023}. It is possible to reexpress $n_\text{ph}$ in terms of the transformed operators, however, the corrected expression, $n_\text{ph}=\frac 12 \braket{(\Piph^\text{D})^2 + (\Qph^\text{D} + \pol\cdot\sum_k \vb{r}_k/\sqrt{\Omega})^2-1}$, contains two-body electronic operators and is comparatively challenging to evaluate.

Instead of calculating these expressions directly, we here propose an alternative approach, based on the Hellman-Feynman theorem. It can be shown (see \cref{app:nph_derivation}) that, both in the Coulomb gauge and the Dipole gauge,
\begin{equation}
\label{eq:nph}
n_\text{ph} = \qty(\frac{|\pol|}{2\Omega} \frac{\partial}{\partial|\pol|} + \frac{\partial}{\partial\Omega}) E.
\end{equation}
In fact, as $E$ is gauge invariant in the complete basis set limit, the same relation can be used in any gauge. Note that evaluation of \cref{eq:nph} requires only a single directional derivative, which can be evaluated either using finite differences (FD) or automatic differentiation (AD). We employ FD for QED-CCSD-2 and AD for AFQMC, following the approach described in Ref.~\onlinecite{mahajan_response_2023} with forward-mode AD. Both calculations take into account the orbital relaxation of the trial or reference wavefunction. 

\begin{figure}
\begin{center}
\includegraphics{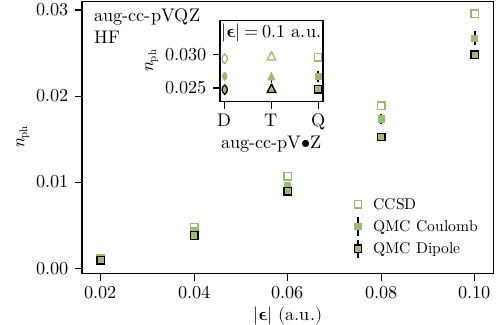}
\end{center}
\caption{The photon number of the HF molecule as a function of the light-matter coupling $|\pol|$ for the molecule HF, using the derivative in \cref{eq:nph} in AFQMC (both gauges) and CCSD (dipole gauge). The inset shows the basis set convergence for a fixed light-matter coupling.}
\label{fig:nph}
\end{figure}

Comparing QED-CCSD-2 calculations in the dipole gauge and AFQMC calculations in both the Coulomb and dipole gauge (\cref{fig:nph}), we find qualitative agreement. However, there are quantitative differences between the three approaches at strong light-matter couplings of the order of 10\%, with the AFQMC dipole gauge photon number being consistently lower and the QED-CCSD-2 photon number being consistently higher than the AFQMC Coulomb gauge result. Since the approaches individually seem converged to the complete basis set limit (\cref{fig:nph} inset), the difference is likely not a gauge dependence caused by the basis-set truncation. Instead, they should arise from the energy inaccuracies in the respective numerical methods, discussed above. In addition, the AD results for AFQMC do not take into account population control, which was found to induce only a comparatively small bias in Ref.~\onlinecite{mahajan_response_2023}. We further note that the statistical fluctuations of the photon number are significantly smaller in the dipole gauge.

\section{Conclusion}
We have presented QED-AFQMC, a 
generalization of the AFQMC algorithm
that can handle a light-matter coupling to a cavity photon mode, both in the Coulomb and in the dipole gauge. We compared AFQMC calculations for small cavity-coupled molecules to both FCI and CCSD.

While finding good agreement with FCI, our calculations show that the accuracy of QED-CCSD can be significantly improved by adding the purely electronic perturbative triples correction. Based on the resulting agreement, the dependence on the remaining correlation pieces on the light matter coupling is expected to be weak.

We further demonstrated that, for the molecules we studied, agreement between the results in the Coulomb and dipole gauges can be reached based on extrapolation of the numerically accessible correlation-consistent Gaussian basis sets.

To calculate photon occupation numbers in a gauge invariant way, based on our multi-gauge multi-method calculations, we propose a Hellman-Feynman theorem-based relation, which can be easily evaluated in any gauge, either using finite differences or automatic differentiation. The resulting photon numbers show some remaining quantitative differences between CCSD and AFQMC in the different gauges. This disagreement may be remedied by further refining the two methods, e.g. by adding the full QED triples correction to CCSD or using more sophisticated trial wave functions in AFQMC.

Beyond the current work, the gauge-flexibility of our AFQMC method also opens the possibility to simulate extended systems, such as cavity-coupled materials, which have been challenging for dipole-gauge calculations and the previously available correlated precision methods. Apart from adding more matter degrees of freedom, a further step to generalize our work would be the addition of multiple cavity modes, which, based on previous applications to electron-phonon systems, is not expected to significantly increase the computational cost of the method.

Therefore, we believe that the QED-AFQMC algorithm will become a versatile addition to a growing numerical toolbox that can help unravel the cavity QED matter problem.
\section{Acknowledgments}
We thank Brandon Kyle Eskridge, Johannes Flick, and Ryan Levy for helpful discussions. Our AFQMC code uses the Carlo.jl~\cite{weber_carlo_2024} framework. The figures were created using Makie.jl~\cite{danisch_makie_2021}. L.W. acknowledges support by the Deutsche Forschungsgemeinschaft (DFG, German Research Foundation) through grant WE 7176-1-1. The Flatiron Institute is a division of the Simons Foundation. 

The data and scripts used to produce the figures of this paper are available in~\onlinecite{data}.
\appendix
\section{Dipole-dipole interaction with or without quadrupole}
\label{app:quadrupole}
In a finite basis set, the dipole-dipole interaction term in $H^\text{D}_2$ from \cref{eq:dipolegauge} can be written either as $(P\vb dP)^2$ or as $P\vb d^2P$, where $\vb{d}=\sum_i\vb r_i$ and $P$ is the projector to the finite basis set.
\begin{figure}[h]
\begin{center}
\includegraphics{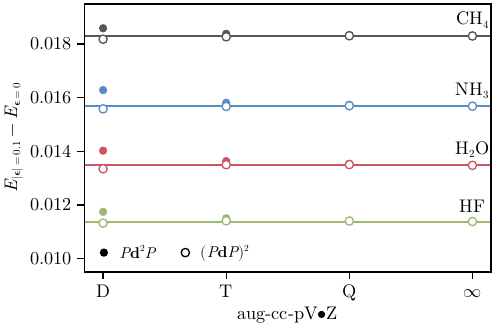}
\end{center}
\caption{The energy difference between $|\pol|=0.1\,\text{a.u.}$ and $|\pol| = 0$ for different molecules, computed using CCSD in different basis sets. The energies with closed symbols were computed with the exact quadrupole matrix elements, $P\vb d^2 P$, while the open symbols were computed with the square of truncated dipoles, $(P\vb d P)^2$. The same matrix elements were used both in the reference and the correlated parts of the calculation. The rightmost values were extrapolated to the complete basis set limit using the aug-cc-pVTZ and aug-cc-pVQZ bases, as in the main text. Horizontal lines where drawn through the extrapolated energies as a guide to the eye.}
\label{fig:quadrupole}
\end{figure}
\Cref{fig:quadrupole} shows that, while the complete basis set limit is insensitive to this difference, the convergence of finite basis set is generally improved by the choice $(P\vb d P)^2$. We therefore perform all calculations in the main text with this choice.
\section{Method comparison without Coulomb repulsion}
\label{app:nonint}
\begin{figure}
\begin{center}
\includegraphics{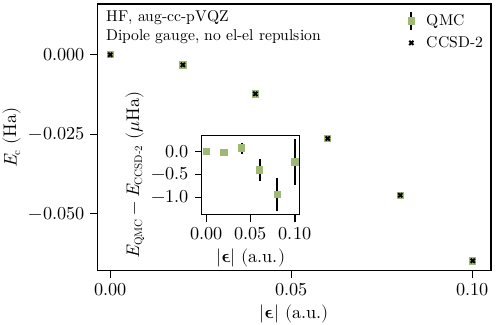}
\end{center}
\caption{The correlation energy of the molecule HF without Coulomb repulsion as a function of the light matter coupling $\pol$, computed using QED-AFQMC and QED-CC, both in the dipole gauge. The inset shows the direct energy difference between the methods.}
\label{fig:nonint}
\end{figure}

In this appendix, we compare the QED-AFQMC and QED-CC methods for the HF molecule without Coulomb repulsion (\cref{fig:nonint}). Although both methods are still approximate in this case, we find excellent agreement within the statistical uncertainty. This suggests that the disagreement found between the methods in the main text originates from the electronic correlations, rather than the electron-photon correlations per-se.

It further stands to reason that rather than including higher-order photon excitations, the most effective way to improve the present QED-CCSD-2 method is to include the triples correction (T), which in first-order approximation, we have done by neglecting its light-matter coupling dependence, (T\textsubscript{0}).
\section{QED-CCSD-2 vs QED-CCSD-1}
\label{app:ccsd1}
\begin{figure}
\begin{center}
\includegraphics{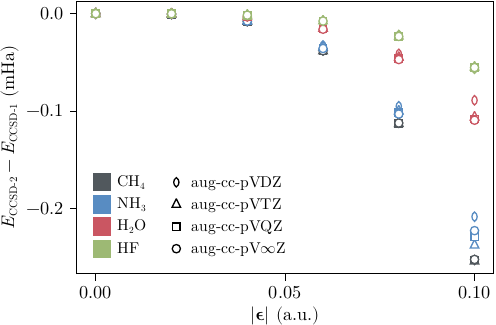}
\end{center}
\caption{The energy difference between QED-CCSD-2 and QED-CCSD-1 for all molecules and basis sets considered in the main text.}
\label{fig:ccsd1}
\end{figure}

We also performed calculations in the QED-CCSD-1 scheme, which only includes single-photon excitations. Comparing to QED-CCSD-2 (\cref{fig:ccsd1}) for all molecules and basis sets considered in the main text, we find that the additional energy contribution upon including double-photon excitations is coupling dependent.
For the strongest light-matter couplings, the energy difference is on the order of 0.1 mHa. This is still much smaller than the difference between QED-CCSD(T\textsubscript{0})-2 and QED-AFQMC found in the main text. Therefore, including further photon excitations without additional electronic excitations, i.e. QED-CCSD-3, is unlikely to account for the remaining disparities.

\section{Derivation of the differential photon number estimator}
\label{app:nph_derivation}
In this appendix, we derive the photon number estimator of \cref{eq:nph} explicitly in both the Coulomb and the dipole gauge and show that they match in the complete basis set limit.

Using the Hellmann-Feynman theorem, we can write
\begin{align}
    n_\text{ph}^{C/D} &= \qty(\frac{|\pol|}{2\Omega} \frac{\partial}{\partial|\pol|} + \frac{\partial}{\partial\Omega}) E^{C/D} \nonumber \\
    &= \Braket{\qty(\frac{\vert\pol\vert}{2\Omega} \frac{\partial}{\partial\vert\pol\vert} + \frac{\partial}{\partial\Omega}) H^{C/D}},
\end{align}
where the expectation value is with respect to the exact ground state wavefunction and the $|\pol|$ derivative is to be understood in the sense of $\partial \pol/\partial |\pol| = \pol/|\pol|$.

For $H^C$, based on \cref{eq:first_quant}, we obtain the expected
\begin{align}
    \label{eq:nph_coulomb}
    n_\text{ph}^C &= \frac 12 \Braket{\Piph^2 + \Qph^2 - 1}.
\end{align}
For $H^D$, based on \cref{eq:dipolegauge}, 
\begin{align}
    \frac{|\pol|}{2\Omega} \frac{\partial}{\partial |\pol|} H^D &=\frac{\Qph^D\pol}{2\sqrt{\Omega}}   \cdot \sum_i \vb{r}_i + \frac{1}{2\Omega} \qty(\pol \cdot \sum_i \vb r_i)^2,\nonumber\\
    \frac{\partial}{\partial\Omega}H_D &= \frac{\Qph^D \pol}{2\sqrt{\Omega}}   \cdot \sum_i \vb{r}_i + \frac{1}{2} \qty((\Piph^D)^2 + (\Qph^D)^2 -1),
\end{align}
where we have explicitly labeled the dipole gauge operators $\Qph^D$ and $\Piph^D$ to distinguish them from the Coulomb gauge. Combined, we have
\begin{align}
n_\text{ph}^D &= \frac 12 \Braket{(\Piph^D)^2 + \qty(\Qph^D+\frac{\pol}{2\sqrt{\Omega}} \cdot \sum_i \vb{r}_i)^2 -1}
\end{align}
To connect back to the Coulomb gauge, we first need to undo both the canonical transformation that switches $\Qph^D \leftrightarrow \Piph^D$ and the PZW transformation, leading to
$\Piph^D = \Qph$ and
\begin{align}
    \label{eq:qph_relation}
    \Qph^D &= U_\text{PZW}^\dagger \Piph U_\text{PZW}\nonumber\\
    &= \Piph - \frac{\pol}{2\sqrt{\Omega}} \cdot \sum_i \vb{r}_i.
\end{align}
So that $n_\text{ph}^C = n_\text{ph}^D$.

This result is not surprising in the complete basis set as the energy $E$ itself and consequently its derivatives are invariant under the unitary PZW transformation.
In a finite basis set, the above derivation still holds formally, but the Hamiltonians $H^C$ and $H^D$ are no longer related by the exact $U_\text{PZW}$ so that \cref{eq:qph_relation} only holds approximately for observables computed using the respective basis-set-truncated Hamiltonians. As with the energy, these discrepancies should vanish in a basis set extrapolation, but may do so at a different convergence rate. In practice, we observe this convergence to be comparatively quick (see inset of \cref{fig:nph}).

\bibliographystyle{achemso}
\def\bibsection{\section*{\refname}} 
\bibliography{paper.bib}

\clearpage
\begin{figure*}
\textbf{\Large For Table of Contents Use Only.}\bigskip

\includegraphics{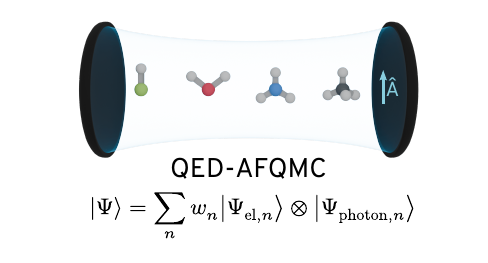}
\end{figure*}

\end{document}